# Relevant Window based Buffer-map Compression in P2P Streaming Media System

Chunxi Li, Changjia Chen, Dahming Chiu

*Abstract*—Popular peer to peer streaming media systems such as PPLive and UUSee rely on periodic buffer-map exchange between peers for proper operation. The buffer-map exchange contains redundant information which causes non-negligible overhead. In this paper we present a theoretical framework to study how the overhead can be lowered. Differentiating from the traditional data compression approach, we do not treat each buffer-map as an isolated data block, but consider the correlations between the sequentially exchanged buffer-maps. Under this framework, two buffer-map compression schemes are proposed and the correctness of the schemes is proved mathematically. Moreover, we derive the theoretical limit of compression gain based on probability theory and information theory. Based on the system parameters of UUSee (a popular P2P streaming platform), our simulations show that the buffer-map sizes are reduced by 86% and 90% (from 456 bits down to only 66 bits and 46 bits) respectively after applying our schemes. Furthermore, by combining with the traditional compression methods (on individual blocks), the sizes are decreased by 91% and 95% (to 42 bits and 24 bits) respectively. Our study provides a guideline for developing practical compression algorithms.

*Index Terms*—buffer-map, bitmap, P2P, compression, relevant window

## I. INTRODUCTION

In recent years, the increasingly development of P2P streaming media system has been attracting more and more attention of researchers. In the majority of the public research reports, people have tried to made use of the discovered *buffer-map message* (BM) to carry out systematic observations and analysis in different aspects, such as buffer descriptions [2] -[9], the startup performance [3][4][4], peer offset and offset lag [5], and data fetching strategies[12][13][14]. However, no one has paid attention to BM itself.

Generally speaking, BM is designed to depict the buffer filling states of a peer, and it is periodically exchanged between paired peers for informing each other which data the other can and can't share. The BM exchange between peers produces some non-neglectable overhead and that is becoming an important cause for concern. According to our measurement on some top popular P2P streaming media systems including PPLive and UUSee, the BM overhead for one peer is at least about *30kbps* and *8kbps* respectively. This overhead may not reduce with the decreasing of the video playback rate, such as in a narrowband wireless environment. Moreover, it makes things worse in the case of encountering unstable network conditions, because a peer needs to connect with much more peers for escaping the bad situation. Besides that, potential overhead increase can also result from the findings [1] that decreasing the time period of BM exchange can help streaming content diffusion. On the other hand, our measurement based analysis shows there is much redundant information need to be removed from BM because the filling state of each piece of data will be repeatedly reported many times to a receiver peer. Therefore, no matter in which circumstances, it is necessary to decrease the BM overhead.

Obviously, data compression is the most operational way. Traditional lossless data compression methods [16]-[22] can be applied to reduce the BM size. In fact, in 2008 we do first found a type of combining algorithm of Lempel–Ziv (LZ) [17][18] and run-length encoding (RLE) [19] is adopted in UUSee through our measurement study, which decreases the buffer-map from 456 bits to 140 bits, and later we got to know that PPLive adopted certain 2-level algorithms of Huffman [16] algorithms to do that. All the traditional methods treat each BM as a general and independent data piece. However, the successive exchanged BMs are strongly correlated with more than half of the information in a BM being redundant. E.g., a peer needs not to report a chunk buffer state in its BM if the peer on the other side has downloaded that chunk. Due to not recognizing the characteristic of BM exchange, much redundant information still remains in the buffer-map compressed by traditional method. How to remove the redundancy information as far as possible? How to evaluate the compression efficiency? What are the fundamental limits of the size of the compressed buffer-map? All such issues are very interesting to both researchers and system designers. To the best of our knowledge, the study on buffer-map compression has never been reported in either practical or theoretical level, not to mention the serious and thorough study.

Chunxi Li. Author is with the Electronic and Information Engineering College, Beijing Jiaotong University, No.3 Shang Yuan Cun,Hai Dian District Beijing,China Post-Code 100044 (phone: 86-01-51684759 ext. 115; fax: 86-01-51683682; e-mail: chxli1@ bjtu.edu.cn).

Changjia Chen Author, is with the Electronic and Information Engineering College, Beijing Jiaotong University, No.3 Shang Yuan Cun,Hai Dian District Beijing,China Post-Code 100044 (e-mail: changjiachen@sina.com).

Dahming Chiu Author is with Department of Information Engineering, the Chinese University of Hong Kong, Ho Sin Hang, Room 836, Hong Kong, China (e-mail: dmchiu@ie.cuhk.edu.hk).

In this paper, we present an original and bilateral framework which opens another door for BM compression. Totally different from the traditional data compression principle, we don't treat each buffer-map as a general and independent data block, but recognize the correlations between the sequential exchanged buffer-maps so as to exclude majority redundant information from a regular BM. Moreover, our approach does not conflict with traditional data compression principle but can work together with it. Our contributions in this paper include: *i)*. we present a fundamental framework of compression based on two crucial but easily overlooked compression principles discovered from BM exchange; *ii)*. Under the original framework, we present two efficient BM compression schemes, the feasibilities of which are proved from mathematics viewpoint; *iii)*. The theoretical limit of average size of the compressed BM is deduced based on probability theory and information theory. The numerical results according to UUSee's system parameters show that, if without transmission errors, the buffer-map can be reduced by 86% and 90% from 456 bits down to only 66 bits and 46 bits respectively according to the two schemes we presented; Furthermore, if combining with the traditional data compression principle, it can be decreased by 91% and 95% to 42 bits and 24 bits respectively.

In the remaining of this paper, we highlight the importance and contributions of our research in the context of an overview and related work about P2P streaming media system in section II; Our core idea and the fundamental theoretical framework are presented in section III; Section IV puts forward two BM compression schemes under the theoretical framework and proofs the feasibility theoretically; In section V, we in-depth analyze the theoretical limit of the compressed buffer-map based on probability theory and information theory, and discuss the simulation results with UUSee's system parameter. Section VI concludes the paper.

## II. AN OVERVIEW OF P2P STREAMING MEDIA SYSTEM

Referring to Fig.1, a typical P2P streaming media system uses few servers to serve large number of audiences (named as *peer*) with both live and VoD programs by sharing the capacities of all the individuals as a whole [6]-[10].

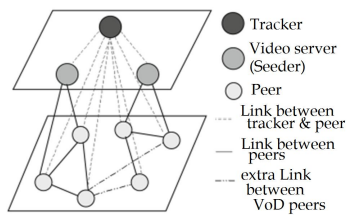

Fig. 1.. The system structure

In such a system, the seeder will divide the media streaming into continuous data blocks called *chunk*, and inject them into the network according to peers' requirement. Each chunk has a unique ID which is sequentially assigned in ascending order, i.e. the earlier played chunk has the smaller ID. In the other side, each peer will use a buffer organized with chunk units as Fig.2 shown, to cache the chunks received from other peers in most cases or the seeder in few cases for smooth playback and more significantly, sharing with other peers. Live peer only caches a small fraction of the whole video, while VoD peer may cache almost the whole video. Peer's buffer is usually partially filled due to the influence of many factors. The downloaded chunks (the shadow square in Fig.2) can be shared, while the empty areas need to be filled by downloading from others. This is the sharing principle playing the key role in the similar P2P content distribution system.

For enabling the key sharing principle between P2P users, a *buffer-map message* (*BM*) is introduced to exchange the buffer information between the paired peers. Referring to Fig.2, a peer's BM contains two parts, an *offset* $\varphi$ and a *bitmap* $b$. The *offset* $\varphi$ corresponds to the oldest chunk, i.e., the smallest chunk ID in the buffer, and the bitmap is a $\{0,1\}$ sequence $b=(b_0,...,b_{|b|-1})$, which represents the buffer filling state. The length $|b|$ indicates the buffer size. In bitmap $b$, a bit value 1(0) at the $i^{th}$ ($0 \leq i < |b|$) component $b_i$ means that the peer has(has not) the chunk with $ID_{\varphi+i-1}$ in its buffer. For simplicity, we call such a buffer-map message as a *regular* BM($\varphi$, $b$).

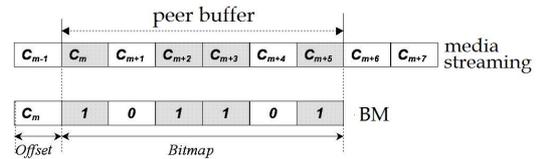

Fig. 2. Buffer and buffer-map

According to our measurement on the top popular P2P stream media systems including PPLive and UUSee, a peer sends out a BM of *250-byte* or *80-byte* long every *4* or *5 seconds* respectively, and each peer keeps at least *30 connections* with other peers concurrently. Therefore, the BM overhead for one peer is at least about *30kbps* and *8kbps* respectively. On the other hand, since a peer constantly removes the ever played chunk from its buffer by right shifting the buffer head pointer, and fetches new chunks to fill the buffer, the chunk IDs at both ends of the buffer will move forward with time. This whole progress is reflected in the periodically reporting BMs. However, since BM exchange time period is much shorter than the buffer size (measured in playback duration), the state of each unique chunk will be repeatedly reported by many sequential BMs. E.g., in general a UUSee peer has a buffer 140s long, and sends BM every 5s, thus each chunk will be roughly reported up to 28 times on average. Obviously, there is too much redundant information in a regular BM.

The performance directly connects to the BM exchange time period [1][24][25]. Since the faster buffer-map exchange can leads to the smaller initial buffering delay, but also the excessive overhead, each P2P streaming media system has an overhead-delay tradeoff. For a given overhead constraint, an efficient compression on buffer-map means a performance improvement in terms of initial buffer delay and the ability to overcome flush crowd in reverse proportion to compression ratio according to theorem 3 and corollary 1 in [1].

Traditional lossless data compression methods can be applied to reduce the size of buffer-map. However, they are incapable of removing the redundant information between successive exchanged buffer-maps. In this paper, we seek to establish a new theoretical framework for buffer-map compression, which belongs to a totally different system other than that of traditional data compression approaches. The new compression theory can guide us to devise the efficient and even the most powerful practical compression algorithms if combing traditional data compression principle.

### III. THE BASICS IDEALS AND CORE CONCEPT

As we know, most redundant BM information is ascribed to the large number of repeat state reports of the same buffer positions in bitmap. More specifically, once a buffer position is filled, it will be filled forever and needs not to be repeatedly reported in later buffer-maps. In the other side, only those buffer positions with 0-value in current *bitmap* may change their values to '1' in the subsequent bitmaps. Such a seemingly trivial observation on the buffer-map exchange leads to a nontrivial compression insight: it is not necessary to include all the buffer positions in a bitmap.

To facilitate figurative understanding the basic ideas, we use two simple analogies to illustrate our points.

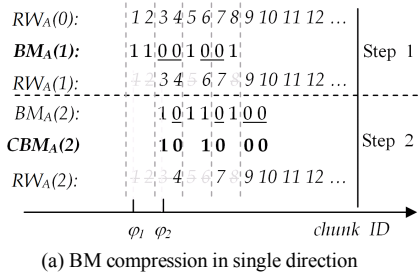

(a) BM compression in single direction

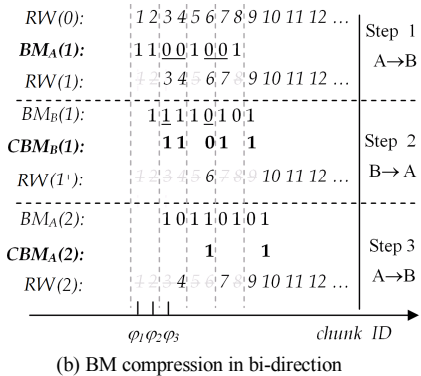

(b) BM compression in bi-direction

Fig. 3. The demonstrations of BM compression

For simplicity, we assuming an ideal communication situation without packet loss and transmission delay. Supposing a buffer-map exchange progress from peer A to peer B according to this idea as shown in Fig.3(a), $BM_A(i)$ means the $i^{th}$ BM of peer A to be sent to peer B, and RW will be explained later. In the first step, after sending the $BM_A(1)$, peer B gets to know which chunks can be downloaded from peer A. Thus, for $BM_A(2)$, peer A only needs to report those positions which have value 0 in $BM_A(1)$ but excludes those with value 1s in $BM_A(1)$, as well as the new positions at $\{9,10\}$. As a result, we get the compressed BM as $CBM_A(2)$ which will be sent to peer B. Furthermore, once a chunk is fetched, a peer will never care if other peers have the same chunk or not. The exchange sequence based on the ideas is shown in Fig.3(b). In the first step, after sending the $BM_A(1)$, peer B gets to know which chunks can be downloaded from peer A, as well as which chunks are needed by peer A; When $BM_B(1)$ is produced, peer B only extracts those positions peer A needs, i.e. the positions which are either value 0 or never announced in $BM_A(1)$ but excludes those with value 1s in $BM_A(1)$. As a result, peer B sends the compressed BM $CBM_B(1)$ to peer A; In the beginning of step 3, peer A already knows which of its downloaded chunks have ever been told to peer B and which chunks peer B doesn't cared. At last, by removing those positions with determinate state of value 1 from $BM_A(2)$, we obtain a $CBM_A(2)$ only 2 bits long.

Base on these observations, we conclude two fix??(exclusion) principles on how to compact the buffer-map.

*Principle 1*: A peer never needs to report a buffer position further in his bitmap once sending a value 1 in this position.

*Principle 2:* A peer never needs to report a buffer position to a receiver peer further once receiving a value 1 in this position from that receiver peer.

It can be seen from the above illustration, unlike a regular buffer-map, the compact bitmap itself cannot fix its locations. Only by mapped to the right position sequence, can the compressed bitmap be correctly encoded and decoded. Thus, we introduce the concept of *relevant window* to describe those positions which are not excluded by above principles. In Fig.3, the RW is just short for the conception. Generally speaking, a *relevant window* (RW) is a set of unique nonnegative integers arranged in ascending order, each of which element corresponds to an ID of chunk which has never been reported with value 1 in the buffer-map. Since a *relevant window* has infinite elements in theory, mathematically, we express it as $RW\ L = \{l_0 < l_1 < \cdots < l_N\}$ and assume $\forall l > l_N \Rightarrow l \in L$ in above expression. By the way, the maximum excluded position (MEP) for a given *relevant window* $L$ is defined as $mep(L) = \max\{p \notin L : p < l_N\}$. The *relevant window* can be interpreted as all the chunks which have not been downloaded, while the MEP means the largest *chunk-id* which has ever been fetched.

The update sequence of *relevant window* in the sender is shown in Fig.3. Comparing to this progress, it is not difficult to deduce the update sequence in the receiver. When sending a buffer-map, the sender first encodes the bitmap according to its RW, and then removes those positions which have state value 1 from its RW. When a compact buffer-map arrives, the receiver first decodes it according to the corresponding RW, and then deletes those positions from RW if the sender reports them with state value 1. Thus, based on the illustration in Fig.3, if without data transmission error and delay, and assuming the same initial RWs in both sides, the RWs in both sides are always synchronous. Therefore, in the new compression mechanism, similar to the bitmap in a regular BM, the compressed bitmap has the same format of a {0,1} sequence as $v=(v_0,...,v_{|v|-1})$, while the difference lies on that the value of $v_i$ is explained as the filling state on the buffer position $l_i$ in the *relevant window L*.

In summary, the principles discovered from BM exchange lays a solid foundation for us to establish a *relevant window based* framework for buffer-map compression. The overhead for BM exchange can be significantly reduced if we can use buffer-map on *relevant window* instead of on the regular buffer window.

## IV. THE BM COMPRESSION BASED ON RELEVANT WINDOW

Given the same *relevant window L*, the sender peer can uniquely extract the compact states from the regular buffer-map and the receiver peer can losslessly decode the compact bitmap. The key points in the BM compression based on *relevant window* are *i)* how to look for the proper *relevant window L*, and *ii)* how to keep the consistency of *relevant windows* between sender and receiver.

In light of the exclusion principles, the relevant windows based BM compression can be applied into either a single peer or a pair of peers. If using principle 1 alone to construct the *relevant window*, we have the BM compression scheme based on single peer's *relevant window* (BMCS_SRW); If both principles are considered, we have the BM compression scheme based on paired peers' common *relevant window* (BMCS_CRW). The latter is a little more complex than the former but it has much better compression efficiency. In the following theoretical discussions, we assume an ideal network situation without packet loss and transmission error.

Moreover, because these schemes are essential different from traditional methods like Huffman, LZ and RL, there are some rooms to invent certain algorithms to make use of the joint forces. E.g., traditional methods can reinforce the compression after our methods and vice versa.

### A. BM compression scheme based on single peer's relevant window

Considering a peer, say peer *A*, is sending the compact buffer filling states $v(t)$ based on its relevant window $L_A(t)$ to its neighbor peers at time *t* under *principle* 1, for correctly encoding and decoding the bitmap $v(t)$ in the sender A and receiver B respectively, certain exchange mechanism must be designed to ensure the consistency of the *relevant windows* in both sides at any time. For that, it intuitively requires each paired connection should keep the communication from the very beginning in theory, while in practice such a requirement can be met by either periodically announcing the compressed BM or by reporting the current relevant window. After that, the relevant windows will be correctly and sequentially updated according to each BM exchange. Strictly speaking, following lemma gives the definite answer.

*Lemma 1*: In the BM sending sequence from one peer to another according to the BM compression scheme based on single peer's relevant window (BMCS_SRW), the consistency of both the *relevant windows* in peer A for encoding and in peer B for decoding, as well as the correctness of the decoded buffer-map corresponding to the encoded buffer-map, can be fully assured.

*Proof*: Let's consider a process of BMCS_SRW including a sending protocol and a receiving protocol.

In the sending side, assuming at time *t*, peer *A* has a *relevant window* $L_A(t)=\{l_0<l_1<...<l_N\}$. For a given *regular* BM ($\varphi$, $b=(b_0,...,b_{|b|-1})$), the bitmap $b$ can be divided into two subsets: the unwanted bits which have ever been reported with value 1s and no longer need to be sent out, and the wanted bits including all the rest bits. In other words, we can get the wanted bits subset $v$ by extracting the bitmap $v=\{b_{l-\varphi}: l \in L_A(t), 0 \leq l-\varphi <|b|\}$ from regular bitmap $b$ according to the positions listed in *relevant window* $L_A(t)$. Thus, the bitmap $b$ can be expressed as

$$b_k = \begin{cases} v_i, & k = l_i - \varphi \\ 1, & L_A(t) - \varphi \neq k \text{ and } k \leq |b| \end{cases} \quad (1)$$

In this equation, the bit sequence of $v_i$ is just the compact bitmap in the compressed BM $CBM(\varphi, v)$. After finishing the BM compressing, *relevant window* $L_A(t)$ will be updated to $L_A(t^+)$ by eliminating the positions from the $L_A(t)$ on condition that either $l<\varphi$ or $b_{l-\varphi}=1$.

In receiving side, for correctively decoding the received compressed buffer-map, peer *B* should hold a same *relevant window* (we call it $L_{B/A}$) as peer A's $L_A$. When receiving a $CBM(\varphi, v)$, peer *B* first updates the current *relevant window* $L_{B/A}(t)$ to $L'_{B/A}(t)=\{l'_0<l'_1<...<l'_N\}$ by removing those chunks which no longer exist in peer A, i.e. the positions meeting the condition $l_i<\varphi$ out of $L_{B/A}(t)$. Then, we can decode the bit sequence $v=(v_0,...,v_{|v|-1})$ by mapping it onto the positions $\{l_i: \varphi \leq l'_0<l'_1...<l'_{|v|-2}<l'_{|v|-1}\}$ in $L'_{B/A}(t)$ where $|v|$ is the length of sequence $v$.

For entirely recovering the uncompressed bitmap $b'$, we first calculate the length of vector $b'$ as

$$|b'| = max(mep(L'_{B/A}(t)), l'_{|v|-1}) - \varphi, \quad (2)$$

where $mep(L) = max\{p \notin L : p < l_N\}$ is the maximum excluded position for a given relevant window *L*. Then the we have a recovered bitmap $b'$ as

$$b'_k = \begin{cases} v_i, & k = l'_i - \varphi \\ 1, & L_{B/A}(t) - \varphi \neq k \text{ and } k \leq |b'| \end{cases} \quad (3)$$

After the decoding, relevant window $L'_{B/A}(t)$ will be updated to $L_{B/A}(t^+)$ by deleting the positions which meet the condition the filling states $v_i=1$ out of $L'_{B/A}(t)$.

According to the above encoding and decoding sequence, if given the condition that the relevant window $L_{B/A}(t)$ in peer $B$ equals the relevant window $L_A(t)$ in peer $A$, i.e., $L_{B/A}(t)= L_A(t)$, we have *i)* the recovered bitmap $b'$ is the same as the original bitmap $b$, i.e., $b=b'$, and *ii)* the *relevant window* $L_{B/A}(t^+)$ in peer $B$ equals the relevant window $L_A(t^+)$ of peer $A$. Thus, lemma 1 is proved. ∎

*Corollary 1*: If a peer plays the video steadily and continually, it is not necessary to include the field of offset $\varphi$ in the compressed BM for synchronizing the relevant windows in both sides.

*Proof:* The entire unfilled chunk positions in peer $A$ are depicted by its *relevant window* and each element in the RW will be sequentially ruled out if the chunk corresponds to the element is timely downloaded before its playing time. We know *offset* $\varphi$ indicates the current playback location. Therefore, a smooth playback in peer $A$ means each position $l_i$ in its relevant window must satisfy the condition $\varphi \leq l_i$ at any time t. Under this condition, using the similar process of proof to *lemma 1*, it is easy to derive the corollary 1. ∎

This corollary guides to a more efficient BMCS_SRW by removing the offset in some of the compressed BM. Occasionally, some few chunks may be not timely downloaded and missed out for playing in a real situation. In that cases, the peer $A$ needs to contain an field like offset in the compressed BM sending to peer $B$ in order to let peer $B$ adjust its *relevant window* $L_{B/A}$ for keeping the consistency. Another purpose of including the offset in the compressed BM is to declare which range of chunks is really buffered in the sender peer. However, memory price today is so cheap that a client may a large buffer to cache nearly the whole video content, thus in certain sense, the buffer is unlimited. Therefore, we no more need to use the offset to locate the available buffer range.

At last, in the optimized BMCS_SRW, we adapt two types of compressed BM to improve the compression ratio. It should be noted that we adopt the same shorten symbol of *CMB* here but with a little different meaning as before. For a given regular BM ($\varphi, b= (b_0, ..., b_{|b|-1})$) to be send at time *t*, peer $A$ will send out a *CBM (v)* of type 0 if $\varphi \leq l_0$ and *CBM (k, v)* of type 1 if $l_0 < l_1 < ... < l_{k-1} < \varphi \leq l_k$, where $\{l_0, l_1, ..., l_{k-1}, l_k\} \subseteq L_A(t)$. In both types of BM, the sequence $v$ can be expressed as $v=\{b_{l-\varphi}: l \in L_A(t), 0 \leq l-\varphi < |b|\}$. After sending out the BM, peer $A$ will get new relevant window $L_A(t^+)$ by deleting those positions meeting the condition of either $l<\varphi$ or $b_{l-\varphi} =1$ from its relevant window $L_A(t)$.

From the operable point of view, the consistent relevant windows of $L_A(t)$ in peer $A$ and $L_{B/A}(t)$ in peer B are the most import to correctly encoding and decoding the buffer-map. One primary task for peer $B$ is to accurately update the *relevant window* based on the received compressed BM. We omit the decompressing progress of sequence $v$ due to its simpleness. If receiving a type 0 *CBM (v)* at time $t$, peer B will delete each position $l_i$ satisfying the condition $v_i=1$ from its relevant window $L_{B/A}(t)$ and get a new relevant window $L_{B/A}(t^+)$. If receiving a type 1 *CBM (k, v)* at time $t$, peer B will first obtain new relevant window $L'_{B/A}(t)$ by removing $k$ smallest positions $\{l_0, l_1,..., l_{k-1}\}$ from $L_{B/A}(t)$, and then delete all the positions meeting the condition $v_i=1$ out of the $L'_{B/A}(t)$ for getting the final relevant window $L_{B/A}(t^+)$.

*B. BM compression scheme based on paired peer's common relevant window*

According to the scheme discussed above, a peer will keep a set of relevant windows which are independent from one another. One of them is used to compress the peer's buffer-map for sending, and all others as used to decode the received compressed buffer-map. Because in the encoding process of BMCS_SRW a peer compresses the buffer-map only based on its own relevant window but regardless those of its neighbor peers, the impacted BM may include many unwanted positions which are already filled in the buffer of receiver peer. Therefore, there is much room to further improve the compression ratio.

By combining *principle 2*, all the unwanted positions for the receiver peer can be ruled out based on a better designed *relevant window*. Moreover, in that *relevant window*, both *principles 1* and *2* can be enforced jointly by the peers who exchange BM. In this enforcement, the paired peers, say peer $A$ and peer $B$, seek to collectively maintain a *common relevant window* (*CRW*) consisting of those positions which are never announced with a filling state value *1s* by the both sides. With the same *common relevant window* in both paired peers, a BM compression scheme (BMCS_CRW) can be described as follows.

We assume the paired peers $A$ and $B$ hold a same common relevant window. For the convenience of discussion, we denote it as $L_{AB}$ and $L_{BA}$ in peer $A$ and peer $B$ respectively. Assuming the BM is exchanged every $T$ seconds in each direction, and peer B will send its BM $\tau$ seconds later than peer A. In other words, peer $A$ has a regular BM ($\varphi, b= (b_0, ..., b_{|b|-1})$) to be sent at the time $t=iT$ and peer $B$ has a regular BM ($\omega, d= (d_0, ..., d_{|d|-1})$) to be sent at the time $t=\tau +iT$, for i=0,1,…and $0<\tau \leq T$.

At time $t=iT$, peer $A$ applies the sending protocol to get and send out the *CBM* ($\varphi, v$), where the sequence $v=\{b_{l-\varphi}: l \in L_{AB}(t), 0 \leq l-\varphi <|b|\}$. The relevant window is then updated to $L_{AB}(t^+)$ by deleting those positions from $L_{AB}(t)$ under conditions either $l<\varphi$ or $b_{l-\varphi}=1$. Once receiving a compressed BM ($\varphi, v$) from peer $A$, peer B will call function of receiving protocol. Peer B's relevant window will be first updated to a new one $L'_{BA}(t)$ by removing all the positions less than $\varphi$ from relevant window $L_{BA}(t)$. Then the bit sequence in $v$ can be mapped onto the $|v|$ smallest positions $\{l_0, l_1, ..., l_{|v|-1}\}$ in the relevant window $L'_{BA}(t)$. The length of recovered bitmap vector $\boldsymbol{b}'$ corresponds to that of peer $A$ can be estimated as

$$|\boldsymbol{b}'| = max(mep(L'_{BA}(t)), l_{|v|-1}) - \varphi, \qquad (4)$$

Then peer $B$ can recovers $\boldsymbol{b}'$ as

$$b'_k = \begin{cases} v_i, & k = l_i - \varphi \\ 1, & L_{BA}(t) - \varphi \ne k \text{ and } k \le |\boldsymbol{b}'| \end{cases} \qquad (5)$$

After decoding the BM, peer $B$'s relevant window will be updated to $L_{BA}(t^+)$ by removing those positions $l_i$ meeting condition $v_i=1$ out of $L'_{BA}(t)$.

In the reverse direction, when peer $B$ reports to peer $A$ at time $t = \tau + iT$, peer $B$ will apply the same procedure of sending protocol as peer $A$ to encode and send out the *CBM* ($\omega$, $\boldsymbol{u}$), where the sequence $\boldsymbol{u} = \{d_{l-\omega}: l \in L_{BA}(t), 0 \le l-\omega < |\boldsymbol{d}|\}$. Since all those positions ever reported with value 1 by peer $A$ have already been ruled out in $L_{BA}(t)$, the *CBM* ($\omega$, $\boldsymbol{u}$) satisfies both *principles* 1 and 2. Next, peer B will update its relevant window to $L_{BA}(t^+)$ by deleting those positions from $L_{BA}(t)$ under conditions either $l<\omega$ or $d_{l-\omega}=1$.

On receiving the *CMB*($\omega$, $\boldsymbol{u}$) from peer $B$, peer A will apply the same receiving protocol as peer B. Peer A will first update its *relevant window* to a new one $L'_{AB}(t)$ by removing all the positions less than $\omega$ from $L_{AB}(t)$. Then, the bit sequence in $\boldsymbol{u}$ will be mapped into the $|\boldsymbol{u}|$ smallest positions $\{l_0, l_1, ..., l_{|u|-1}\}$ in relevant window $L'_{BA}(t)$. The length of recovered bitmap vector $\boldsymbol{d}'$ corresponds to that of peer $B$ can be estimated as

$$|\boldsymbol{d}'| = max(mep(L'_{AB}(t)), l_{|u|-1}) - \omega, \qquad (6)$$

Peer $A$ then recovers $\boldsymbol{d}'$ as

$$d'_k = \begin{cases} v_i, & k = l_i - \omega \\ 1, & L_{AB}(t) - \omega \ne k \text{ and } k \le |\boldsymbol{d}'| \end{cases} \qquad (7)$$

After decoding the BM message, peer $A$ will update its *relevant window* to a new one $L_{AB}(t^+)$ by deleting those positions $l_i$ satisfying condition $u_i=1$ from $L'_{AB}(t)$.

According to the encoding and decoding sequence, if given the initial condition that the common relevant window $L_{AB}(t)$ in peer A equals the relevant window $L_{BA}(t)$ in peer B, i.e., $L_{AB}(t) = L_{BA}(t)$, we can draw the conclusions *i)* Although the recovered bitmap $\boldsymbol{d}'$ or $\boldsymbol{b}'$ may be not totally the same as the original one $\boldsymbol{d}$ or $\boldsymbol{b}$, all the wanted filling states are fully contained in $\boldsymbol{d}'$ or $\boldsymbol{b}'$ corresponding to $\boldsymbol{d}$ or $\boldsymbol{b}$ respectively, while all the possible inconsistent states in $\boldsymbol{b}'$ or $\boldsymbol{d}'$ are unwanted and not cared by the receiver peers B or A; *ii)* The *relevant window* $L_{AB}(t^+)$ of peer A always equals the *relevant window* $L_{BA}(t^+)$ of peer B. ∎

Therefore, we conclude above discussion as following lemma 2.

*Lemma 2:* For a pair of peers who exchange buffer-maps according to BMCS_CRW, if given the initial condition that the relevant window $L_{AB}(t_0)$ in peer A equals the relevant window $L_{BA}(t_0)$ in peer B, i.e., $L_{AB}(t_0) = L_{BA}(t_0)$, the consistency of the *common relevant windows* $L_{AB}(t)$ in peer A and $L_{BA}(t)$ in peer B, as well as the consistent representation of all the wanted buffer filling states in the exchanged BM can be fully assured.

### C. Some issues in theory and in practical level

In the above discussions, we assume an ideal system without transmission error and delay. Although the actual conditions are far from the assumption, one may devise some kinds of engineering algorithm efficient enough to approach the theoretical limit. We will further the researches including the engineering design, the additional overhead evaluation and the complexity assessment in the future studies. However, we would like to discuss a few fundamental practical issues in this paper.

There are some basic design philosophies. Firstly, according to the theoretical analysis, certain type of reliable BM exchange protocol needs to be developed to make sure both peers (who exchange BM) have exactly the same understanding of what is received or not. Secondly, the lost buffer-map message needs not to be retransmitted. It will be too later for the retransmission because the time to deducing a packet loss is much longer than the BM sending time period. Thirdly, the buffer-map should be sent out evenly and sequentially corresponding to its production cycle.

For one thing, we only need a quite simple reliable protocol which only provides certain type of confirmation mechanism for the transmitted buffer-map message but without including retransmission of the lost packet. Many reliable transmission approaches can be applied. One way is to trigger an acknowledgement (ACK) for each sending BM, e.g., the receiver can attach the ACK to its next sending BM instead of using a specific ACK protocol message. The ACK itself can be expressed as a very short relative offset but not a 4-byte long offset. Another way is to trigger a negative acknowledgement (NAK) if BM loss occurs. The message number of NAK may be far less than that of ACK, while the loss deduction is far more time consuming.

For another thing, it is unacceptable for a peer to execute a next sending process by waiting the confirmation of the latest sending buffer-map. Once a buffer-map is produced, the sender should send it out immediately after compressing according to the *relevant window*. In engineering, the *relevant window* can be directly derived from an ever sending buffer-map according to BMCS_SRW, or from two buffer-maps, the ever sending one and the latest received one according to BMCS_CRW. There are different design choices for generating the *relevant window* for encoding and decoding. For example, in a sending protocol of BMCS_CRW, one way to generate the *relevant window* is based on the latest received buffer-map and the latest ever sending buffer-map regardless of the confirmation, another way is according to the latest received buffer-map and the latest ever sending buffer-map confirmed by the receiver. For the former, a smaller size of compression BM can be expected because the *relevant window* may be relative newer than the latter for without considering the confirmation in the former. However, due to packet loss and network delay, it is easy for a receiver of the former to get stuck, i.e., because the receiver may get a compressed BM but can't find its corresponding *relevant window*, the BM exchange process has to stop. The Latter has no such question. Just as the regular BM exchange for the Latter, if compressed BM is lost, it is not a big deal since the up-to-date information is not available only briefly; a moment later, another copy of the BM will be sent.

According to both our schemes, certain method should be designed to keep the *relevant window* in a receiver consistent to the sender's. One most direct solution is to include the indications of both the latest (confirmed) buffer-maps in the compressed buffer-map message. The ACK in our simple reliable protocol counts as one indicator of the receiver's buffer-map, the other indicator should be the relative offset of the latest ever sending buffer-map (confirmed by the receiver).

As a conclusion, in a simple protocol design, two extra indicators should be added in the compressed buffer-map message and they help to maintain a consistent relevant window on both sides. The size of both the indicator should be designed as smaller as possible, e.g. 8-bit is a good and actual choice.

## V. THE LENGTH ESTIMATION OF COMPRESSED BITMAP

In this section, we will discuss the theoretical limit values of compressed bitmap in different schemes, including our newly presented schemes, as well as the schemes by combining BMCS_SRW and traditional method, and combining BMCS_CRW and traditional method. We study the bitmap length of the new schemes we present by probability and statistics approach, and study the bitmap lengths of the joint force schemes by information theory.

### A. Preliminary Assumptions

We assume a CBR video with the playback rate $r$ in the following discussion. Under this assumption, the service curve $s(t)=rt$ [3] is the most advanced chunk ID that has been fed into the system. For any stable peer $q$, the *playback delay* $N_q(t)$ is defined as the difference between the service curve $s(t)$ and its *offset* $\varphi_q(t)$, i.e., $N_q(t)=s(t)-\varphi_q(t)$. Since each peer plays the video with the same playback rate $r$, the playback delay is time-independent, i.e. $N_q(t)=N_q$. Based on our measurement, a peer $q$ can buffer all the chunks in the scope of $[\varphi_q(t), s(t)]$ at any given time $t$. Thus, the buffer width of a peer is roughly its playback delay (measured in number of chunks). Measurement results show that the playback delays (or say the buffer width) of all peers are roughly the same. Hence, for simplicity in mathematics, we assume all peers have the same buffer width $N$, in other words, the produced bitmap length is $|b|=N$.

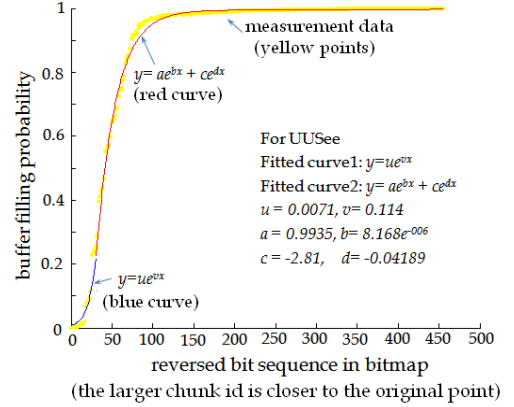

Figure 4. The diffusion S curve in UUSee

Based on our previous research [2], the buffer filling progress in a P2P streaming media system is approximated to be a stationary process. Based on our measurements in April 2009, we draw the buffer filling probability distributions function of UUSee in Fig.4, and the measurement result displays certain S shapes. We fit the S shape by the two-segment curves as curve1 and curver2. The horizontal axis is the reversed local buffer position, i.e., the larger buffer position is nearer the original point with a smaller filling probability. The vertical axis is the buffer filling probability. A point $S(x)$ on the S curve indicates the filling probability at the buffer position $N-x-1$, $0 \le x \le N-1$, in other words, the $i^{th}$ bit $b_i$ in the bitmap $b$ takes a value 1 with the probability $S(N-1-i)$.

In this paper we use $Z(c,t)$ to represent the event that the given chunk $c$ has not been fetched at a given time $t$ in a peer. The probability of the event $Z(c,t)$ can be easily derived from the S curve as:

$$p(Z(c,t)) = \begin{cases} 0, & if \quad c < st - N \\ 1 - S(s_t - c), if & s_t - N \le c \le s_t \\ 1, & if \quad c > s_t \end{cases} \quad (8)$$

where $s_t$ is the service curve, i.e. the largest chunk-id in the system at time $t$.

### B. Bitmap Length of BMCS_SRW

*Theorem 1*: In BMCS_SRW, the size of compressed bitmap is totally determined by the diffusion S curve. Specifically, if given the diffusion function $S(x)$, we have the average size of compressed bitmap $W_{SRW}$

$$W_{SRW\_1} = N - \sum_{i=0}^{N-rT-1} S(i), \quad (9)$$

when containing the offset field in a BM, or

$$W_{SRW\_2} = rT + N - \sum_{i=0}^{N-1} S(i), \quad (10)$$

when without containing the offset field in a BM.

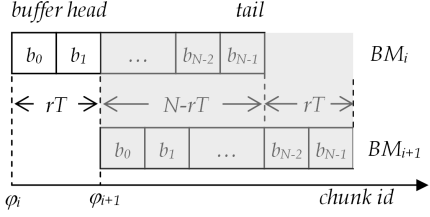

Fig.5. BM Exchange model for BMCS_SRW

*Proof:* As shown in Fig.5, we assume at each time $t=iT$, $i=0,1,\ldots$, peer A announces its compressed buffer-map $CBM(\varphi_i, v_i)$ and updates its relevant window to $L_i$ when finishing the compressing. In other words, $L_i$ is the *relevant window* for next BM compression in peer A at time $t=(i+1)T$. According to principle 1, $L_i$ only includes those chunk positions which are either never announced yet or ever announced a value 0 at time $t = iT$. Thus, we have

$$L_i = \{c: c > \varphi_i + N \text{ or } Z(c, iT)\} \quad (11)$$

Assume peer A has an offset $\varphi_i$ at the time $t = iT$, then

$$\varphi_{i+1} = \varphi_i + rT. \quad (12)$$

If containing the offset $\varphi_{i+1}$ in the compressed buffer map, the average compressed bitmap length is

$$W_{SRW\_1} = rT + \sum_{c=\varphi_i+1}^{\varphi_i+N-1} p(Z(c,iT))$$
$$= rT + (N-rT) - \sum_{c=\varphi_i+1}^{\varphi_i+N-1} S(\varphi_i + N-1-c) = N - \sum_{i=0}^{N-rT-1} S(i) \quad (13)$$

If not including the offset $\varphi_{i+1}$ in the compressed buffer map, the average compressed bitmap length is

$$W_{SRW\_2} = rT + \sum_{c=0}^{\varphi_i+N-1} p(Z(c,iT))$$
$$= rT + N - \sum_{c=0}^{\varphi_i+N-1} S(\varphi_i + N-1-c)) = rT + N - \sum_{i=0}^{N-1} S(i) \quad (14)$$

∎

The compressed bitmap length of the latter is a little larger than the former. However, for the former, a complete buffer map message needs to include the *offset* field which can be designed to a relative *offset* $W_\varphi$ occupying several few bits ($W_\varphi \leq 8$) in most cases. According to the optimized implementation of BMCS_SRW using both two types of BM, the average BM length should be within $[W_2, W_1 + W_\varphi]$.

### C. Bitmap Length of BMCS_CRW

We assume both peer A and peer B exchange their buffer-map to each other. Peer A sends out its buffer-map at time $t = iT$ and $t = \tau + iT$ is the time for peer B to send.

Referring to Fig.6, at time $t = iT$, peer A initially announces its compressed buffer map $CBM(\varphi_i, v_i)$ and updates its relevant window to $L'_{AB}(i)$ when finishing the sending. According to principle 1, the relevant window $L'_{AB}(i)$ only includes those chunk positions which are either never announced yet or ever announced a value 0 at time $t = iT$ by peer A. Upon receiving $CBM(\varphi_i, v_i)$, peer B updates its *relevant window* by removing the positions which are less than $\varphi_i$ and the positions which *filling states* equal value 1 in $v_i$ during the decoding process.

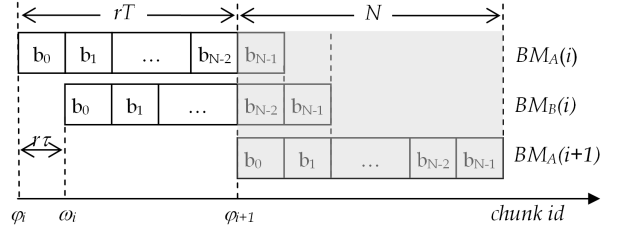

Fig.6. BM Exchange model for BMCS_CRW

Shortly after at time $t = \tau + iT$, peer B sends out its $CBM(\omega_i, u_i)$. Then, peer B will update its relevant window to $L'_{BA}(i)$ by deleting those positions which are less than $\omega_i$ or those positions which filling states is value 1 in $u_i$. Upon receiving BM $(\omega_i, u_i)$, peer A update its *relevant window* to $L_{AB}(i)$ by removing the positions which are less than $\omega_i$ and the positions with *filling states equal to* 1 in $u_i$ out of $L'_{AB}(i)$ during the decoding process. According to principles 1 and 2, $L_{AB}(i)$ only includes those chunk positions which are never reported with value 1 yet by both peer A and B till time $t = \tau + iT$, i.e.,

$$L_{AB}(i) = \{c: c > \max(\varphi_i, \omega_i) + N \text{ or } (Z(c,iT) \text{ and } Z(c, \tau + iT))\} \quad (15)$$

In the next round at time $t = (i+1)T$, the peer A will send a $CBM$ with new offset:

$$\varphi_{i+1} = \varphi_i + rT. \quad (16)$$

To facilitate the analysis, we assuming the same *playback delay* for both paired peers, i.e. they have the same offsets at any time $t$. Under that assumption, we have $\omega_i = \varphi_i + r\tau$. Therefore, the average compressed bitmap length of peer A should be

$$W_{CRW\_AB} = r(T-\tau) + \sum_{c=\varphi_i+N}^{\omega_i+N-1} p(Z(c, \tau + iT)) + \sum_{c=\varphi_{i+1}}^{\varphi_i+N-1} p(Z(c,iT))p(Z(c,\tau+iT))$$
$$= rT - \sum_{c=\varphi_i+N}^{\omega_i+N-1} S(\omega_i + N-1-c))$$
$$+ \sum_{c=\varphi_{i+1}}^{\varphi_i+N-1} ((1-S(\varphi_i+N-1-c))(1-S(\omega_i+N-1-c)))$$
$$= rT - \sum_{i=0}^{r\tau-1} S(i) + \sum_{i=0}^{N-1-rT} ((1-S(i))(1-S(r\tau+i))). \quad (17)$$

Similarly, by substituting the $\tau$ with $T-\tau$, we get the average compressed bitmap length of peer B

$$W_{CRW\_BA} = r\tau - \sum_{i=0}^{r(T-\tau)-1} S(i) + \sum_{i=0}^{N-1-rT} ((1-S(i))(1-S(r(T-\tau)+i))). \quad (18)$$

∎

Thus we proof the following theorem:

*Theorem 2*: In BMCS_CRW, the average size of the compressed bitmap is totally determined by the diffusion S curve. Specifically, if given the diffusion function $S(x)$, under the same playback delay condition for both paired peers, the average bitmap sizes $W_{CRW}$ in each direction are, respectively

$$W_{CRW\_AB} = rT - \sum_{i=0}^{r\tau-1} S(i) + \sum_{i=0}^{N-1-rT}((1-S(i))(1-S(r\tau+i))) \quad (19)$$

and

$$W_{CRW\_BA} = r\tau - \sum_{i=0}^{r(T-\tau)-1} S(i) + \sum_{i=0}^{N-1-rT}((1-S(i))(1-S(r(T-\tau)+i))). \quad (20)$$

where $N$, $T$, $r$ and $\tau$ are the system parameters buffer length, BM exchange time period, video playback rate and the BM sending time interval between the two peers respectively.

Therefore, the average bitmap length over this connection is

$$W_{CRW} = (W_{CRW\_AB} + W_{CRW\_BA})/2. \quad (21)$$

### D. Bitmap Length of Traditional Compression

The idea of traditional lossless compression is deeply connected with statistical inference, and Claude Shannon lays the theoretical foundation [21]. According to information theory [15], the regular bitmap can be regard as the combination of $N$ independent binary sources $\{h_k, 0 \le k \le N-1\}$. As each chunk's downloading can be affected by many factors including network conditions, peer selection and data fetching policy, for peer in stable condition, it is reasonable to assume the $N$ binary sources are independent to each other. For convenience, we use this operator in the following discussions

$$H(x) = x\log_2 x + (1-x)\log_2(1-x). \quad (22)$$

Therefore, with traditional compression, for any regular BM($\varphi$,**b**), we have the compressed bitmap size $W_{Trad}$

$$W_{Trad} = -\sum_{c=\varphi_i}^{\varphi_i+N-1} H(p(Z(c,iT))) = -\sum_{i=0}^{N-1} H(S(i)). \quad (23)$$

### E. Bitmap Length of BMCS_SRW with Traditional Compression

We borrow Fig.5 to explain this derivation process. Obviously, any two binary sources of $h_k$ and $h_{k-rT}$ in the two adjacent $BM_i$ and $BM_{i+1}$ respectively are the same, i.e. both of them correspond to the same chunk. As in BMCS_SRW, if $h_k$ sends value 1 in $BM_i$, then $h_{k-rT}$ in $BM_{i+1}$ must be 1; while if $h_k=0$ in $BM_i$, then $h_{k-rT}$ in $BM_{i+1}$ may be 1 with certain probability. For simplicity, we use symbol $q_{i,j}(c)$, $j=i+1$ to represent the condition probability that the chunk $c$ in $BM_i$ is not downloaded but downloaded at next $BM_{i+1}$. For a stable process of buffer filling, we have following equation:

$$1 - p(Z(c,t_i)) + p(Z(c,t_i))\, q_{i,j}(c) = 1 - p(Z(c,t_j)) \quad (24)$$

Then, we have

$$q_{i,j}(c) = \frac{p(Z(c,t_i)) - p(Z(c,t_j))}{p(Z(c,t_i))} = \frac{S(s_{t_j}-c) - S(s_{t_i}-c)}{1 - S(s_{t_i}-c)} \quad (25)$$

where $s_{t_j} - N \le c \le s_{t_i}$. Therefore, the theoretical limit of bitmap size $W_{JFS}$ in this joint force scheme is

$$\begin{aligned}W_{JFS} &= -\sum_{c=\varphi_{i+1}+N}^{\varphi_{i+1}+N-1} H(p(Z(c,iT+T))) - \sum_{c=\varphi_{i+1}}^{\varphi_i+N-1} P(Z(c,iT))H(q_{i,j}(c)) \\ &= -\sum_{i=0}^{rT-1} H(S(i)) - \sum_{i=0}^{N-rT-1}(1-S(i))H\!\left(\frac{S(i+rT)-S(i)}{1-S(i)}\right)\end{aligned} \quad (26)$$

∎

Thus we proof the following theorem:

*Theorem 3*: For the scheme with the joint force of BMCS_SRW and traditional compression, if given the chunk diffusion function $S(x)$, under the same playback delay condition for both paired peers, the size of the compressed bitmap has a theoretical limit value

$$W_{JFS} = -\sum_{i=0}^{rT-1} H(S(i)) - \sum_{i=0}^{N-rT-1}(1-S(i))H\!\left(\frac{S(i+rT)-S(i)}{1-S(i)}\right), \quad (27)$$

Where the parameters $N$, $T$ and $r$ are the buffer length, BM exchange time period and video playback rate respectively.

### F. Bitmap Length of BMCS_CRW with Traditional Compression

Let's recall that in BMCS_CRW only the states of those positions which have not been buffered in both paired peers should be reported in a compressed buffer-map. We use Fig.6 to explain this derivation process here. We assume peer $A$ sends its BM to peer $B$ at time $iT$ and peer $B$ sends to peer $A$ at time $\tau + iT$. For peer $A$ at time $t=(i+1)T$, the probability to send the state about a chunk $c$ is $p(Z(c,iT)) \times p(Z(c,iT+\tau))$; according to (25), the condition probability that chunk $c$ is not downloaded at time $t=iT$ but downloaded at time $t=(i+1)T$ is $q_{i,i+1}(c)$.

Therefore, in the direction of peer $A$ to peer $B$, the theoretical limit $W_{JFC\_AB}$ of the compressed bitmap size in this joint force is

$$\begin{aligned}W_{JFC\_AB} &= -\sum_{c=\varphi_{i+1}}^{\varphi_{i+1}+N-1} H(p(Z(c,iT+T))) \\ &\quad -\sum_{c=\varphi_i+N}^{\varphi_i+N-1} p(Z(c,iT+\tau))H(p(Z(c,iT+T))) \\ &\quad -\sum_{c=\varphi_{i+1}}^{\varphi_i+N-1} p(Z(c,iT))p(Z(c,iT+\tau))H(q_{i,i+1}(c)) \\ &= -\sum_{i=0}^{r(T-\tau)-1} H(S(i)) - \sum_{i=0}^{r\tau-1}(1-S(i))H(S(i+rT-r\tau)) \\ &\quad -\sum_{i=0}^{N-rT-1}(1-S(i))(1-S(i+r\tau))H\!\left(\frac{S(i+rT)-S(i)}{1-S(i)}\right).\end{aligned} \quad (28)$$

Clearly, the result consists of three parts: the newly increased part, the overlapped parts of both $BM_A(i+1)$ and $BM_B(i)$, and the overlapped part of the three BMs ($BM_A(i+1)$, $BM_B(i)$ and $BM_A(i)$).

Correspondingly, by substituting the $\tau$ with $T-\tau$, we get the theoretical limit $W_{JFC\_BA}$ of compressed bitmap sized in the reverse direction from peer $B$ to peer $A$

$$W_{JFC\_BA} = -\sum_{i=0}^{r\tau-1} H(S(i)) - \sum_{i=0}^{r(T-\tau)-1}(1-S(i))H(S(i+r\tau)) \\ -\sum_{i=0}^{N-rT-1}(1-S(i))(1-S(i+rT-r\tau))H\left(\frac{S(i+rT)-S(i)}{1-S(i)}\right) \quad (29)$$

Therefore, the average bitmap length over this connection between peer A and peer B is

$$W_{JFC} = (W_{JFC\_AB} + W_{JFC\_BA})/2 \quad (30)$$
∎

Thus we proof the following theorem:

*Theorem 4*: For the scheme with the joint force of BMCS_CRW and traditional compression, if given the chunk diffusion function $S(x)$, under the same playback delay condition for both paired peers, the average size of the compressed bitmap has a theoretical limit,

$$W_{JFC} = -\frac{1}{2}\begin{pmatrix} \sum_{i=0}^{r(T-\tau)-1} H(S(i)) + \sum_{i=0}^{r\tau-1}(1-S(i))H(S(i+rT-r\tau)) \\ + \sum_{i=0}^{N-rT-1}(1-S(i))(1-S(i+r\tau))H\left(\frac{S(i+rT)-S(i)}{1-S(i)}\right) \\ + \sum_{i=0}^{r\tau-1} H(S(i)) + \sum_{i=0}^{r(T-\tau)-1}(1-S(i))H(S(i+r\tau)) \\ + \sum_{i=0}^{N-rT-1}(1-S(i))(1-S(i+rT-r\tau))H\left(\frac{S(i+rT)-S(i)}{1-S(i)}\right) \end{pmatrix}, \quad (31)$$

where $N$, $T$, $r$ and $\tau$ are the system parameters of buffer length, BM exchange time period, video playback rate and the BM sending time interval between paired peers respectively.

### G. Simulation with UUSee

By substituting the real parameters of UUSee, i.e. $T$=5s, $r$=3.37 chunks/s, $N$=456 bits, and $S(x)$ as shown in Fig.4, we calculate the average bitmap lengths of all the compression schemes, including the single peer's *relevant widow* based scheme (BMCS_SRW), the *common relevant widow* based scheme (BMCS_CRW), the joint force scheme (BMCS _JFS) of combing BMCS_SRW with the traditional method, and the joint force scheme (BMCS _JFC) of combing BMCS_CRW with traditional method. The numerical results show that in an ideal situation, the bitmap size can be reduced by 86% and 90% from 456 bits down to only 66 bits and 46 bits by BMCS_SRW and BMCS_CRW respectively. Furthermore, by combing the traditional compression approach, the size can be decreased by 91% and 95% to 42 bits and 24 bits respectively. The improvement from BMCS_SRW to BMCS_JFS and from BMCS_CRW to BMCS _JFC is ascribed to the traditional probability inference acting on the new compression theories we present. The detailed numerical results are listed in table I.

By adjusting parameter of the BM exchange period $T$, we figure out the different size of the compressed bitmap. The result is shown in Fig.7 and listed in table I. Even though due to adjusting the exchange period $T$ the real network sharing environment may change to some extent so as to influence the diffusion $S$ curve, we believe our results reflect the overall trends.

Fig.7(a) shows all curves of bitmap size vs exchange period $T$ of these compression schemes. Both curves of BMCS_SRW (the curves with the largest slope) corresponding to (9) and (10) of theorem 1 are nearly overlapping and both curves of BMSC_CRW (the curves with the second largest slope) corresponding to (19) and (20) of theorem 2 are identical to each other under condition $\tau$ =T/2. The two curves on the bottom with triangle and inverse-triangle markers are for BMCS _JFS and BMCS _CFS respectively. It can be seen that the average bitmap size linearly increases with the exchange period $T$ for BMCS_SRW and BMCS_CRW, while the latter can bring more than 30% gain over the former due to considering principle 2 besides principle 1 in the latter scheme. Moreover, the encoded bitmap of either BMCS_SRW or BMCS_CRW can be further compressed because there is different filling probability on each position in the encoded bitmap. Therefore by combining the traditional data compression principle, more than 36% and 46% redundant information can be further ruled out from the encoded bitmap of BMCS_SRW and BMCS_CRW by BMCS _JFS and BMCS _JFC respectively.

As a reference, the bitmap limit of the bitmap compressed by traditional approach is also included in Fig.7(a) and table I. Because it treats each BM as a normal data block without considering the BM exchange feature in P2P system, the theoretical limit is a constant value (77 bits) shown as a horizontal line in Fig.7(a). We note both curves of BMCS_SRW and BMCS_CRW intersect the line of the traditional approach at time about $T$=8s and $T$=18s respectively. On the left of the cross-point, our scheme has smaller bitmap size. Considering the fact that the BM exchange usually has a much high frequency such as 500ms in PPSteam and 4s in PPLive and a faster BM exchange can speed up chunk diffusion, the exchange time $T$ at the cross-point is large enough to indicate the importance of our new compression

TABLE I
THE THEORETICAL LIMIT OF THE BITMAP SIZE IN DEFERENT COMPRESSION SCHEMES

size unit: bit

| Schemes | BM sending period $T$ (s) | | | | |
|---|---|---|---|---|---|
|  | 5 | 10 | 15 | 20 | 25 |
| Traditional compression | 77 | 77 | 77 | 77 | 77 |
| BMCS_SRW | 66 | 83 | 100 | 116 | 133 |
| BMCS_CRW | 46 | 57 | 68 | 78 | 87 |
| Joint force 1 (JFS) | 42 | 52 | 60 | 64 | 66 |
| Joint force 2 (JFC) | 24 | 33 | 40 | 45 | 50 |

Joint force 1 means the scheme with the joint force of BMCS_SRW and traditional compression;
Joint force 2 means the scheme with the joint force of BMCS_CRW and traditional compression;
For both BMCS_CRW and Joint force 2, we use $\tau$=T/2 in calculation

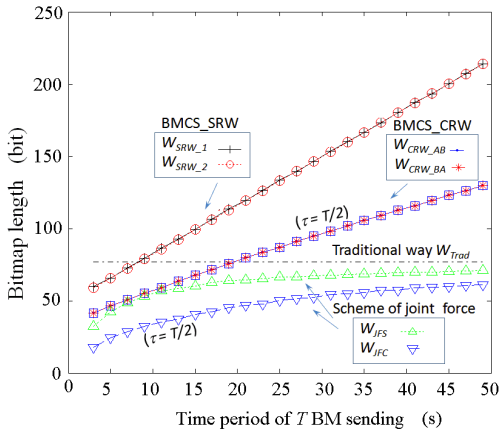

(a) The BM length v.s. period T

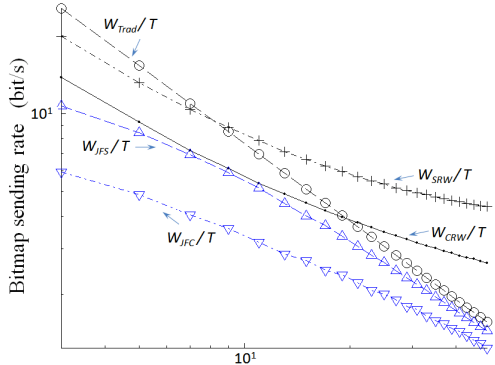

(b) The bit rate v.s. period T

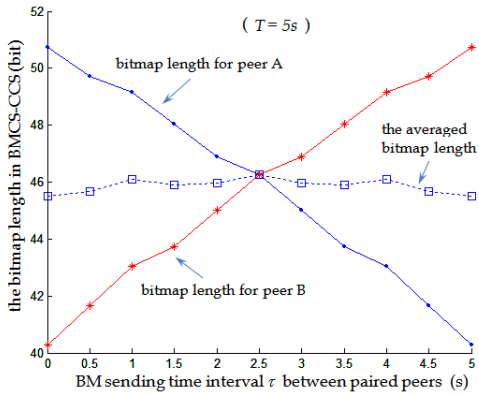

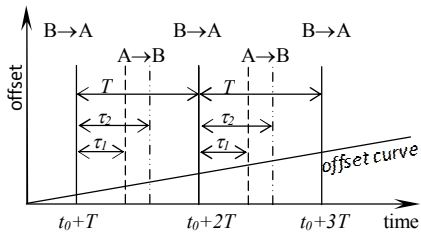

Fig.8. the BM exchange time sequence in BMSC_CRW

(c) the BM length v.s. interval τ

Fig. 7. Numerical results with UUSee approaches. Moreover, because both our new and traditional methods are not mutually exclusive but can work together, by the joint forces, both curves of BMCS _JFS and BMCS _CFS are far below all other curves, i.e., they are much more powerful than any single scheme. In addition, we show the result in another form in Fig.7(b), where the vertical axis is the bit rate of bitmap sending i.e. the bitmap size dividing by exchange time period $T$, and the horizontal axis is $T$. Besides the similar conclusion as drawn from Fig.7(a), we can see that the smaller time period $T$ always leads to the higher bit rate of sending.

Another interesting issue is when to send the BM after receiving one BM in BMSC_CRW is more preferable to make the overall overhead smaller on both directions. Reference to Fig.8, assuming peer B sends out its BM at time $t=t_0+iT$, we need to answer what is the most suitable time $\tau$ behind time $t$ for peer A to send out its BM. For analyzing this problem, we adjust the sending interval $\tau$ within $[0, T]$ while keep the sending period stationary ($T$=5s) in the BM length calculation. The result is shown in Fig.7(c). We can see that the average bitmap size on the both directions is nearly invariable no matter what the sending interval $\tau$ is changed to. It suggests a designer need not to think over the selection of sending interval $\tau$ at all. The same conclusion is also applied to BMCS_JFC.

As a result, the original compression approaches and the theoretical results can guide us to look for powerful engineering design. In fact, based on the basic schemes we present, many efficient practical solutions can be devised to approach the ideal bitmap sizes. However, for the specific engineering designs, many factors need to be considered. For example, because we need both peers (who exchange BM) have exactly the same understanding of what is received or not, certain type of reliable protocol should be designed to implement the BM exchange.

Detailed study on the engineering design issues is beyond the discussion of this paper, and we will focus on these works, including the engineering implementation, the additional overhead evaluation and complexity discussion, in future research works.

## VI. CONCLUSION

In this paper, we present an original theoretical framework for buffer-map message compression in light of the discovered BM exchange principles in P2P system and the introduced important concept of relevant window. Different from the existing general data compression principle, we don't treat each BM as a general and independent data block, but recognize the correlations between the sequential exchanged BMs. In other words, a peer never needs to report a buffer position further in his bitmap once sending a value 1 in this position, and moreover, a peer never needs to report a buffer position to a receiver peer further once receiving a value 1 in this position from that receiver peer. Under the theoretical framework, two efficient buffer-map compression schemes are presented and the feasibility of the schemes is proved in theory. Both the new method we presented and traditional data compression method belong to different theoretical systems in nature, and they don't conflict with each other but can work together. The theoretical sizes of the compressed bitmap for both new schemes we presented as well as the schemes of combining the traditional compression principle are derived in mathematics. At last, the numerical results calculated with system parameters of UUSee validate the efficiency of our methods. The compression ratio is about 14% and 10% for single peer's relevant window based scheme and common relevant window based scheme respectively. Moreover, if combining with the general data compression approach, compression ratios can be further improved to about 9% and 5%.

The most importance of the study in this paper is that we establish a new theoretical framework for buffer-map compression, which is different from traditional data compression theory. The new frame can guide us to devise the efficient engineering solutions, and enlightens us to develop the most powerful solutions by combing traditional data compression principle. We will conduct the study on the algorithms and protocols in engineer in our future research.

## REFERENCES

[1] C. Feng, B.C Li, B. Li. "*Understanding the Performance Gap between Pull-based Mesh Streaming Protocols and Fundamental Limits*", in the Proceedings of IEEE INFOCOM 2009, Rio de Janeiro, Brazil, April 19-25, 2009.

[2] Y.S. Chen, C.J. Chen and C.X. Li, "*Measure and Model P2P Streaming System by Buffer Bitmap*", HPCC'08, Sept. 2008,

[3] C.X Li, C.J. Chen, "*Initial Offset Placement in P2P Live Streaming Systems*", http://arxiv.org/ftp/arxiv/papers/0810/0810.2063.pdf

[4] Zhou, YIPeng Chiu, Dah Ming Chiu, John C.S., "A Simple Model for Analyzing P2P Streaming Protocols", *ICNP 2007*

[5] X.J. Hei, Y.Liu, K. Ross, "Inferring Network-Wide Quality in P2P Live Streaming Systems", *IEEE Journal on Selected Areas in Communications*, vol. 25, no. 10, Dec. 2007.

[6] X. Hei, C. Liang, J. Liang, et al. "A measurement study of a large scale P2P IPTV system", *IEEE Transactions on Multimedia*, 2007, 9(8):1672-1687

[7] X. Hei, C. Liang, J. Liang, Y. Liu, and K. W. Ross, "Insights into PPLive: A measurement study of a large-scale P2P IPTV system," *IPTV workshop in conjunction with WWW2006*, May 2006.

[8] S. Ali, A. Mathur, and H. Zhang, "Measurement of commercial Peer-to-Peer live video streaming," *First Workshop on Recent Advances in Peer-to-Peer Streaming,* Aug. 2006.

[9] L. Vu, I. Gupta, J. Liang, K. Nahrstedt, "MapPing the PPLive Network: Studying the Impacts of Media Streaming on P2P Overlays", *UIUC Tech report*, August 2006

[10] X. Zhang, J. Liu, B. Li, et al. "Coolstreaming/donet: a data-driven overlay network for peer-to-peer live media streaming", *In: Proceedings of IEEE/INFOCOM'05*, Miami, USA, 2005:2102-2111

[11] Y. James Hall, P. Piemonte, M. Weyant, Joost: a measurement study.<*http://www.patrickpiemonte.com/15744-Joost.pdf*>.

[12] A. Vlavianos, M. Iliofotou, and M. Faloutsos, "BiToS: Enhancing BitTorrent for Supporting Streaming Applications", in *Global Internet Workshop in conjunction with IEEE INFOCOM 2006*, April 2006.

[13] S. Tewari and L. Kleinrock, "Analytical Model for BitTorrent-based Live Video Streaming", *Consumer Communications and Networking Conference*, 2007

[14] Zhou, YIPeng Chiu, Dah Ming Lui, John C.S., "A Simple Model for Analyzing P2P Streaming Protocols", *ICNP 2007*

[15] THOMAS M. COVER and JOY A. THOMAS, "Elements of Information Theory", second edition, *published by John Wiley & Sons, Inc. Hoboken, New Jersey*, July 2006, ISBN: 0-471-24195-4, pp 13-33.

[16] D.A. Huffman, "A Method for the Construction of Minimum-Redundancy Codes", *Proceedings of the I.R.E.*, September 1952, pp 1098–1102.

[17] J. Ziv, A. Lempel; "A Universal Algorithm for Sequential Data Compression", *IEEE Transactions on Information Theory*, 23(3), pp.337–343, May 1977

[18] J. Ziv, A. Lempel: "Compression of individual sequences via variable rate coding". *IEEE Transactions on Information Theory*. Vol. IT¨C24, No. 5, Sept. 1978, pp 530¨C535.

[19] Golomb SW, "Run-length encodings", *IEEE Transactions on Information Theory*, 1966 pp 140-149

[20] G.Q Deng, Y.F. Zhang, C.X. Li, C.J. Chen, "*A bitmap coding method for P2P streaming protocols*",in Proc. of CAR, 2010

[21] Rissanen, J.J.; Langdon, G.G. "Arithmetic coding", *IBM Journal of Research and Development* 23 (2): 149–162. March 1979

[22] Ian H.; Neal, Radford M.; Cleary, John G. "Arithmetic Coding for Data Compression". *Communications of the ACM* 30 (6): 520–540, June, 1987

[23] Shannon CE, Weaver W, "The mathematical theory of communication", *University of Illinois Press*, Urbana IL, 1949.

[24] M. Zhang, Q. Zhang, L. Sun, and S. Yang, "Understanding the Power of Pull-based Streaming Protocol: Can We Do Better?" *IEEE J. on Sel. Areas in Communications*, December 2007.

[25] M. Zhang, Y. Xiong, Q. Zhang, and S. Yang, "A Peer-to-Peer Network for Live Media Streaming: Using a Push-Pull Approach," in *Proc. of ACM Multimedia 2005*, November 2005.